\documentclass{article}

\usepackage{microtype}
\usepackage{graphicx}
\usepackage[export]{adjustbox}
\graphicspath{./FIg/}

\usepackage{subfigure}
\usepackage{booktabs} 
\usepackage{amsmath, amssymb, mathtools, amsthm}
\usepackage{hyperref}
\usepackage[capitalize,noabbrev]{cleveref}
\usepackage[accepted]{icml2025}
\usepackage{enumitem}
\usepackage[utf8]{inputenc}
\usepackage{pgfplots}      
\pgfplotsset{compat=1.17}
\usepackage{caption}
\usepackage{subcaption}
\usepackage{multirow}
\usepackage{float}
\usepackage{afterpage}

\theoremstyle{plain}

\theoremstyle{definition}

\icmltitlerunning{TCR2PEP Benchmark}

\begin{document}

\twocolumn[
\icmltitle{TCRTransBench: A Comprehensive Benchmark for Bidirectional TCR-Peptide Sequence Generation}

\icmlsetsymbol{equal}{*}

\begin{icmlauthorlist}
\icmlauthor{Yiming Wang}{westlake,equal}
\icmlauthor{Weiyu Xiao}{hust,equal}
\icmlauthor{Jiangbin Zheng}{westlake,equal}
\icmlauthor{Stan Z. Li}{westlake}
\end{icmlauthorlist}

\icmlaffiliation{westlake}{Westlake University, Hangzhou, China}
\icmlaffiliation{hust}{Huazhong University of Science and Technology, Wuhan, China}

\icmlcorrespondingauthor{Stan Z. Li}{Stan.ZQ.Li@westlake.edu.cn}

\printAffiliationsAndNotice{\icmlEqualContribution}

\vskip 0.3in
]



\begin{abstract} T-cell receptor (TCR) interactions with antigenic peptides underpin adaptive immunity and are pivotal for personalized immunotherapy and vaccine development.Despite recent progress, computational modeling of TCR-peptide specificity remains challenging due to data scarcity, complex sequence dependencies, and the absence of standardized evaluation frameworks.To systematically address these issues, we introduce \textbf{TCRTransBench}, a comprehensive benchmark for bidirectional TCR-peptide sequence generation tasks. Specifically, we define two sequence-to-sequence (seq2seq) tasks: generating antigenic peptides from TCR sequences (TCR2PEP) and generating TCR sequences from antigenic peptides (PEP2TCR). Our framework provides a rigorously curated, MHC-free dataset comprising tens of thousands of validated TCR-peptide pairs, along with diverse evaluation metrics that integrate computational efficiency, sequence accuracy, and biological plausibility. Extensive benchmarking across representative neural architectures—including recurrent, convolutional, and transformer-based models—reveals key trade-offs among performance metrics, highlighting transformers' effectiveness in capturing intricate biological interactions and the necessity of biologically-informed evaluation criteria. TCRTransBench establishes standardized tasks, datasets, and evaluation protocols, laying a robust foundation for future computational advances in immunological sequence modeling and therapeutic protein design.
\end{abstract}


\section{Introduction} T cells are pivotal lymphocytes within the adaptive immune system that recognize infected or abnormal cells through highly specific interactions between their T-cell receptors (TCRs) and antigenic peptides presented by major histocompatibility complex (MHC) molecules on antigen-presenting cells~\cite{annurev.immunol,rossjohn2015t,springer2020prediction}, thereby initiating immune responses. Accurately modeling and predicting TCR-peptide interactions is crucial for advancing personalized immunotherapy, vaccine development, and the engineering of therapeutic proteins~\cite{jurtz2017netmhcpan,lu2021deep,hudson2023can}. Designing antigenic peptides capable of binding specific TCRs, or conversely, engineering novel TCR sequences targeting predefined peptides, represents a transformative yet challenging direction that could significantly expand the repertoire of immunological targets and facilitate controllable immune modulation~\cite{karthikeyan2023conditional,davidsen2019deep,isacchini2021deep,seo2025tcr}.

\begin{figure}
    \centering
    \includegraphics[width=1\linewidth]{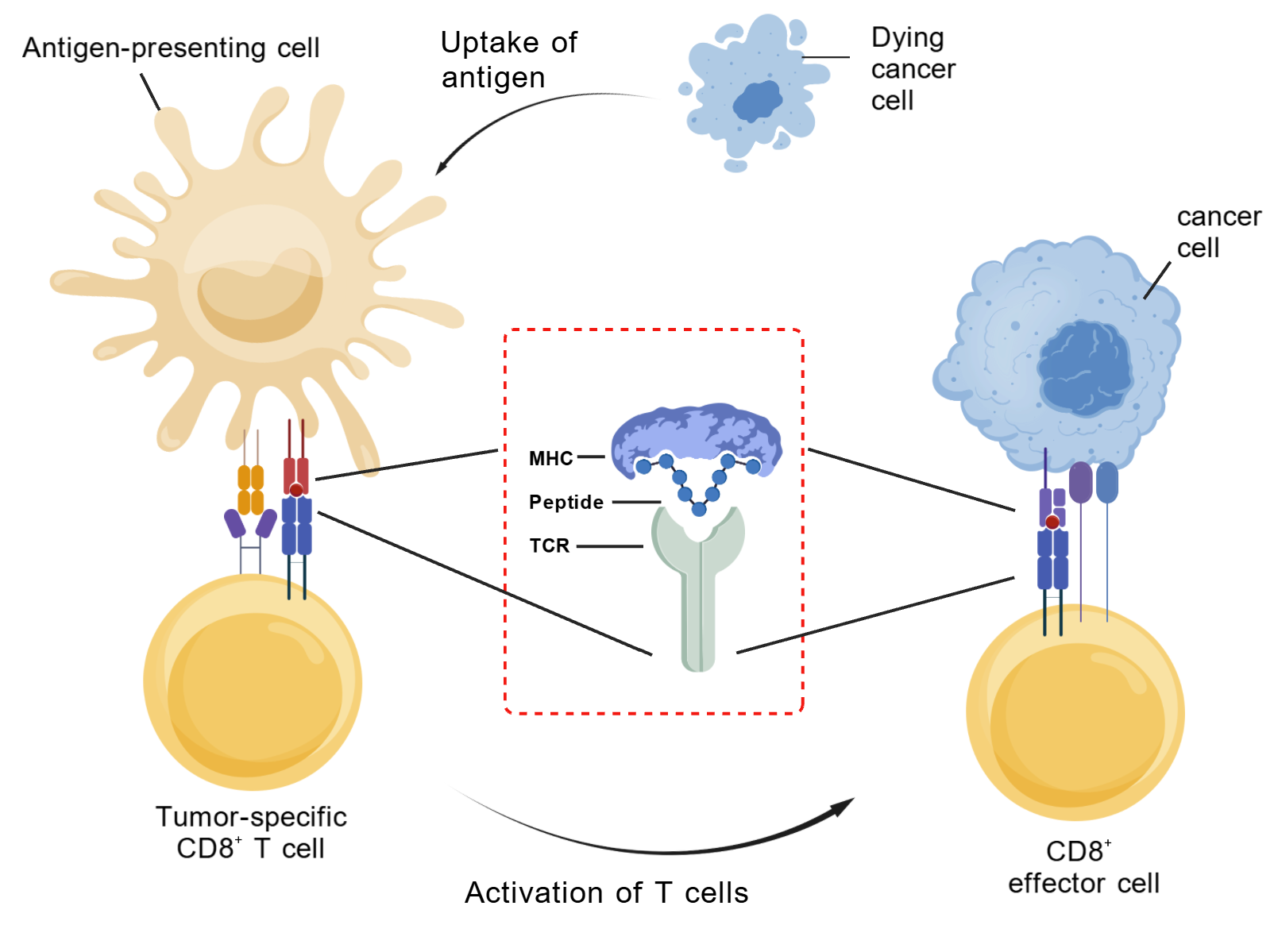}
    \caption{CD8$^{+}$ T-cell--mediated tumor clearance hinges on TCR recognition of cognate peptide--MHC~I: naive CD8$^{+}$ T cells are activated by peptide--MHC~I complexes on antigen-presenting cells and differentiate into effector CD8$^{+}$ T cells; these effectors then recognize cognate peptide--MHC~I on tumor cells via the TCR and precisely induce tumor-cell death.}
    \label{Fig:2}
\end{figure}
Despite its importance, computational prediction of TCR-peptide interactions faces significant challenges. The complexity arises primarily from the highly variable binding regions within TCRs, as well as the inherently many-to-many nature of the TCR-peptide recognition mechanism~\cite{wong2019comparative,birnbaum2014deconstructing,sidhom2021deeptcr,weber2021titan}. Additionally, the limited availability of comprehensive and high-quality paired TCR-peptide datasets constrains model training and validation, making it difficult to achieve generalization to novel, unseen epitopes.\cite{bagaev2020vdjdb} Moreover, existing approaches predominantly model TCR-peptide interactions as discriminative classification tasks, thus lacking the flexibility and generative capability necessary for effective protein design applications~\cite{xu2021dlptcr,springer2020prediction,weber2021titan,weber2024t,qi2025roadmap}.

Recent advances in generative artificial intelligence, particularly sequence-to-sequence (seq2seq) modeling frameworks derived from natural language processing, offer promising solutions to overcome these limitations. Transformer-based seq2seq architectures such as T5~\cite{raffel2020exploring} have demonstrated remarkable effectiveness in capturing complex dependencies within biological sequences~\cite{lin2022language,elnaggar2021prottrans,lin2023evolutionary,watson2023novo}. Although these methods have begun to find applications in protein engineering and design, their specific potential for modeling TCR-peptide interactions remains underexplored. Furthermore, current generative methods either overly rely on known epitopes in their training sets or depend extensively on MHC information~\cite{karthikeyan2023conditional,karthikeyan2025conditional,drost2025benchmarking}, limiting their applicability to scenarios requiring novel sequence generation or zero-shot generalization.

To systematically address these limitations, we propose TCRTransBench, a comprehensive benchmarking framework specifically designed for the bidirectional generation of TCR and antigenic peptide sequences in an MHC-free manner. Our framework formalizes two critical sequence-generation tasks: TCR-to-peptide (TCR2PEP) and peptide-to-TCR (PEP2TCR), employing an autoregressive generative language modeling approach based solely on TCR and peptide sequence representations. Leveraging a rigorously curated dataset containing tens of thousands of validated TCR-peptide pairs from established immunological databases~\cite{tickotsky2017mcpas,shugay2018vdjdb,vita2019immune}, our framework enables robust evaluation of various neural architectures, including recurrent neural networks, convolutional models, and transformer-based architectures.

Moreover, recognizing the limitations inherent in conventional computational metrics, we introduce an integrated evaluation pipeline that combines standard neural machine translation metrics (e.g., BLEU, perplexity, and F1 score) with biologically-informed measures such as predicted binding affinity and sequence novelty. This dual approach ensures that model-generated sequences not only demonstrate computational accuracy but also maintain biological plausibility and relevance, a crucial requirement for practical applications in therapeutic protein engineering.

Our contributions can be summarized as follows: \begin{itemize}[noitemsep, topsep=0pt] \item We propose TCRTransBench, a novel benchmark designed explicitly for the bidirectional generation of TCR and antigenic peptide sequences, providing foundational resources and standardized tasks for immunoinformatics research. \item We perform systematic evaluations of the bidirectional TCR-peptide generation tasks using comprehensive metrics spanning neural machine translation (NMT) measures, model uncertainty assessments, and biologically relevant metrics. \item We establish a set of baseline models representing diverse technical approaches—including recurrent, convolutional, and transformer-based architectures—and train them on our benchmark tasks, thereby providing a solid foundation for future methodological advancements. \item We will publicly release the benchmark dataset and implementation code resources to facilitate further methodological advancements and reproducibility in the computational immunology community in the future. \end{itemize}

\section{Related Work}
\paragraph{Autoregressive Seq2seq Language Model} An autoregressive sequence-to-sequence (seq2seq) language model assigns a probability \( P(w) \) to a sequence \( w = (w_1, w_2, \dots, w_n) \) by factorizing it as the product of conditional probabilities of each token \( w_t \) given its preceding context \( w_1, w_2, \dots, w_{t-1} \), formulated as \( P(w) = \prod_{t=1}^{n} P(w_t \mid w_1, w_2, \dots, w_{t-1}) \). This approach inherently supports sequential modeling tasks by capturing dependencies among tokens. Autoregressive seq2seq models commonly employ architectures that explicitly model both input and output sequences, facilitating tasks involving conditional generation. Typical autoregressive seq2seq models include encoder-decoder architectures, such as T5 \cite{raffel2020exploring}, designed to handle various sequence mapping tasks effectively. These architectures are particularly suitable for our tasks, as they effectively model the bidirectional conditional interactions between TCR and peptide sequences. Additionally, simpler architectures, like 1-dimensional Convolutional Neural Networks (1dCNN), can also serve as autoregressive seq2seq models, offering computational efficiency and the ability to capture local sequence dependencies, albeit typically with less capacity for modeling long-range dependencies compared to transformer-based models.

\paragraph{Deep TCR-Peptide Specificity Prediction}
Recent deep learning approaches for TCR–peptide specificity prediction have leveraged a variety of network architectures and training strategies to capture the intricate interactions between T-cell receptors and antigens. For example, DeepTCR \cite{sidhom2021deeptcr} and ELATE \cite{dvorkin2021autoencoder} utilize deep neural networks and autoencoders to extract latent features from TCR sequences, while NetTCR-2.0 \cite{montemurro2021nettcr} applies convolutional neural networks to combine TCR (primarily CDR3) and peptide sequence information. Models like pMTnet \cite{lu2021deep} and DLpTCR \cite{xu2021dlptcr} employ transfer learning and ensemble techniques to improve prediction across diverse peptides, and ERGO2 \cite{springer2020prediction} and TITAN \cite{weber2021titan}incorporate attention mechanisms and Transformer-based architectures to integrate multi-modal sequence features, thereby advancing the field significantly. While these methods excel in data-rich scenarios, evaluating de novo language model-generated TCR/peptide sequences demands models capable of robust generalization to unseen epitopes without peptide-specific training. We selected Pan-Peptide \cite{gao2023pan}as the benchmark due to its explicit design for zero-shot prediction, a feature critical for assessing novel interactions where experimental data for fine-tuning is inherently absent. This choice prioritizes methodological alignment with the challenges of evaluating synthetic sequences while retaining comparability to existing state-of-the-art approaches.

\paragraph{Generative Protein Design} 
Recent advances in generative artificial intelligence have revolutionized protein design, transforming it from a labor-intensive experimental process into a computational paradigm capable of efficiently exploring vast sequence spaces. Modern AI approaches leverage large-scale protein databases to learn complex, high-dimensional features that capture the intricate relationships between sequence, structure, and function. This has enabled several breakthrough tools: ESM2 \cite{lin2022language} harnesses evolutionary information through language modeling to learn meaningful protein representations, ProteinMPNN \cite{dauparas2022robust} integrates structural constraints to generate physically viable sequences, and RFDiffusion \cite{watson2023novo} employs diffusion models to sample diverse yet functional designs. These computational advances have dramatically accelerated protein engineering, offering new possibilities for therapeutic development and biotechnology applications by combining biological insights with sophisticated machine learning techniques. However, these models primarily serve as general-purpose tools and have not been specifically optimized for modeling TCR-peptide interactions. In contrast, our approach adopts similar foundational principles without relying on their pretrained weights, explicitly focusing on capturing the unique characteristics of TCR-peptide sequence data.

\section{TCRTransBench Framework Design}

\begin{figure*}[h]
    \centering
    \includegraphics[width=1.0\textwidth]{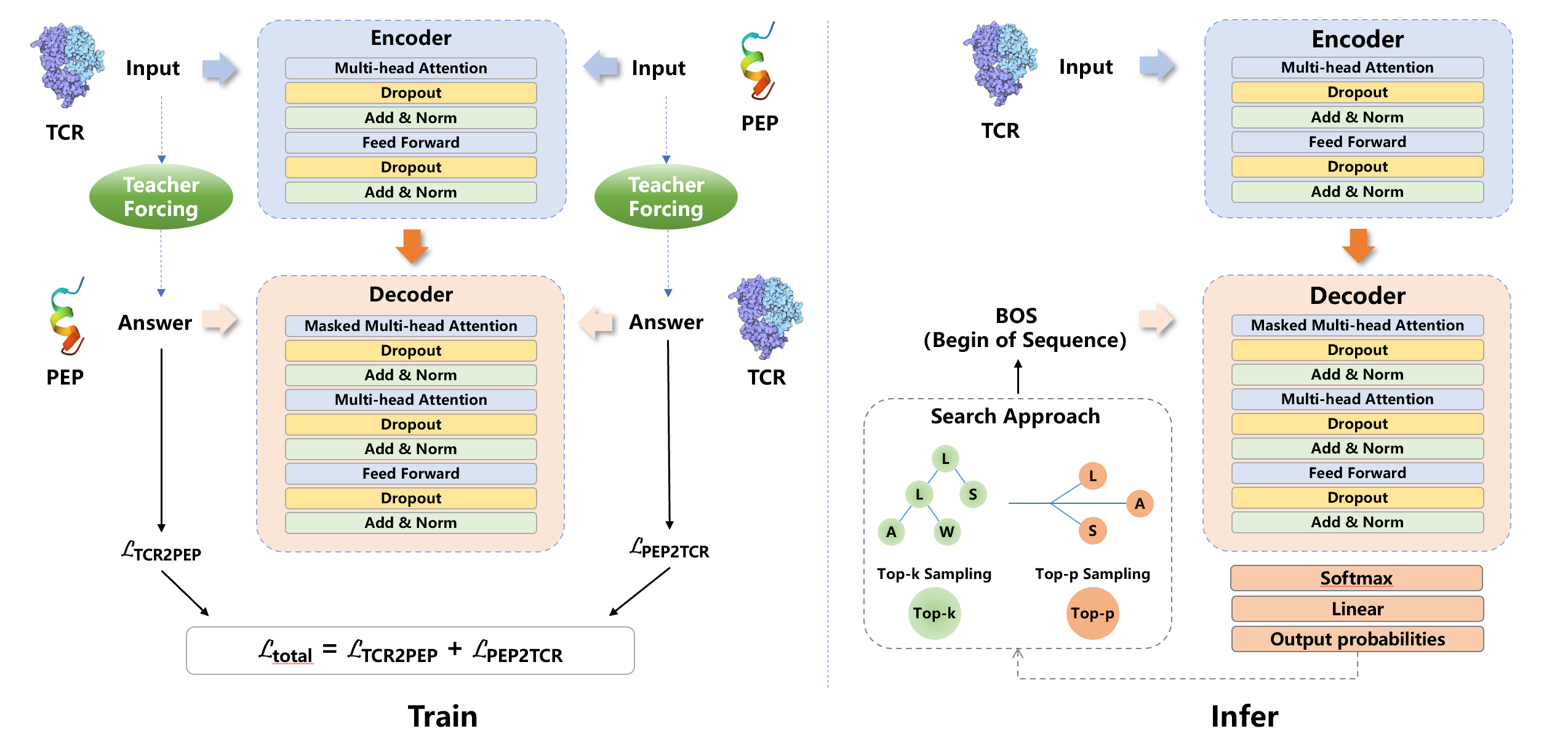} 
    \caption{Training and Inference Strategy}
    \label{fig:2}
\end{figure*}

\subsection{Task Definition}

We consider two complementary tasks, TCR2PEP and PEP2TCR, formulated as bidirectional sequence-to-sequence (seq2seq) modeling problems crucial for computational protein design. \textbf{Unlike existing methods that rely on traditional MHC (Major Histocompatibility Complex) constraints,} these tasks aim to computationally model and design sequences involved in T-cell receptor (TCR) and antigenic peptide interactions. \textbf{This ``MHC-free'' strategy} directly addresses \textbf{a key bottleneck} in personalized immunotherapy and vaccine design.

Formally, given a dataset \(\mathcal{D} = \{(\mathbf{x}^{(n)}, \mathbf{y}^{(n)})\}_{n=1}^{N}\) comprising \(N\) pairs of sequences, we define two key design tasks:
\begin{itemize}[noitemsep]
  \item \textbf{TCR2PEP}: Generating peptide sequences \(\mathbf{y}\) conditioned on known TCR sequences \(\mathbf{x}\), modeled as \(p(\mathbf{y} \mid \mathbf{x})\).
  \item \textbf{PEP2TCR}: Generating TCR sequences \(\mathbf{x}\) conditioned on given peptide sequences \(\mathbf{y}\), modeled as \(p(\mathbf{x} \mid \mathbf{y})\).
\end{itemize}

Here, each T-cell receptor sequence \(\mathbf{x}^{(n)} = [x_1, \ldots, x_{L_x}]\) and antigenic peptide sequence \(\mathbf{y}^{(n)} = [y_1, \ldots, y_{L_y}]\) consist of amino acids from a shared vocabulary \(\mathcal{V}\). To capture their sequential \textbf{biological properties and boundary information}, both sequences are augmented with special start \(\langle \text{START} \rangle\) and end \(\langle \text{END} \rangle\) tokens. This explicit sequential representation enables the models to learn nuanced biochemical interaction patterns essential for accurate sequence prediction.

From the perspective of computational protein design, accurate modeling and generation of TCR-peptide sequences are pivotal. Generating peptides conditioned on specific TCR sequences facilitates targeted therapeutic design, such as personalized vaccines that elicit desired immune responses. Conversely, generating TCR sequences conditioned on known peptides enables the rational design of engineered T-cells with predefined specificity and affinity, significantly enhancing adoptive cell therapy strategies, particularly in cancer treatment and autoimmunity.

Algorithmically, we approach this problem by employing a unified autoregressive language model \(p_\theta\) to capture the \textbf{joint} conditional distributions \(p(\mathbf{y} \mid \mathbf{x})\) and \(p(\mathbf{x} \mid \mathbf{y})\). Leveraging \textbf{the power of} large pre-trained language models, our method benefits from robust feature learning and generalization capabilities, enabling effective capture of \textbf{complex long-range biological dependencies} within sequences. This integration of advanced computational modeling with precise biological constraints represents a significant advancement over traditional structure-based or MHC-dependent methods, paving the way for accelerated, reliable, and broadly applicable computational protein design in immunotherapy.

\subsection{Dataset Construction}
The experimental foundation of TCRTransBench is built upon a rigorously curated dataset of TCR-peptide pairs derived from well-established immunological databases including McPAS~\cite{tickotsky2017mcpas}, VDJdb~\cite{shugay2018vdjdb}, and IEDB~\cite{vita2019immune}. The raw data is subjected to an extensive preprocessing pipeline involving duplicate removal, probabilistic inference of missing fields, and stringent quality filtering. Only high-quality human TCR entries—featuring complete sets of peptide sequences, CDR3$\beta$ sequences, V$\beta$ gene segments, and J$\beta$ gene segments—are retained. This results in a final corpus comprising over 50,000 validated sequence pairs, which are partitioned into training and test sets using a peptide-based stratification strategy. Such a partitioning guarantees that the antigenic peptides in the evaluation set are completely isolated from those in the training set, thereby providing an unbiased assessment of model generalization to unseen TCR-peptide recognition patterns.

\subsection{Evaluation Metrics}
We adopt a comprehensive set of evaluation metrics to \textbf{rigorously and holistically} assess the performance of our MHC-free, bidirectional sequence generation model. The metrics are grouped into three main categories: (i) \textbf{NMT Metrics}, (ii) \textbf{Computational Efficiency}, and (iii) \textbf{Biological Plausibility}.

\paragraph{NMT Metrics}
The NMT metrics consist of two subcategories: \emph{Sequence Quality} and \emph{Model Uncertainty}.

\begin{itemize}[noitemsep]
  \item \textbf{Sequence Quality}: This subcategory evaluates the fidelity and structural integrity of the generated sequences. \textbf{These metrics, originating from machine translation, are used to quantify the \textit{syntactical correctness} of the generated amino acid sequences.}
  \begin{enumerate}[label=(\alph*)]
    \item \textbf{BLEU Score}: Used to evaluate the n-gram precision between generated and reference sequences at the character level.
    
    \item \textbf{Edit Distance}: Measures the minimum number of single-character operations (insertions, deletions, or substitutions) required to transform a generated sequence \(\hat{\mathbf{y}}\) into its reference \(\mathbf{y}\).
    
    \item \textbf{F1 Score}: Computes the harmonic mean of precision and recall at the character (token) level.
  \end{enumerate}
    \item \textbf{Model Uncertainty}: This subcategory quantifies the confidence of the model's predictions using perplexity. A lower perplexity indicates \textbf{a better fit to the data distribution and thus} higher confidence.
\end{itemize}

\paragraph{Computational Efficiency}
Computational efficiency is evaluated through two key metrics:
\begin{itemize}[noitemsep]
  \item \textbf{Inference Speed}: Denoted by \(T\), this metric represents the average time (in seconds) required to generate a single sequence.
  \item \textbf{Model Parameter Count}: Denoted by \(P\), this is the total number of parameters in the model, providing an estimate of its computational complexity and resource requirements.
\end{itemize}

\paragraph{Biological Relevance}
Biological relevance is assessed by evaluating both the binding specificity and the novelty of the generated sequences.
\begin{itemize}[noitemsep]
  \item \textbf{Binding Specificity}: For each generated TCR-peptide pair \((\hat{\mathbf{x}}, \hat{\mathbf{y}})\), binding specificity is quantified using the Pan-Peptide model \cite{gao2023pan}, a \textbf{pre-trained} deep neural network \(f_{\text{PP}}: \mathcal{V}^{L_x} \times \mathcal{V}^{L_y} \to [0,1]\) that outputs a binding probability. Formally, the binding probability is given by
  \[
  p_{\text{bind}}(\hat{\mathbf{x}}, \hat{\mathbf{y}}) = f_{\text{PP}}(\hat{\mathbf{x}}, \hat{\mathbf{y}}; \theta),
  \]
  where \(\theta\) denotes the model parameters. A higher \(p_{\text{bind}}\) indicates a stronger predicted binding specificity. \textbf{Crucially, this metric assesses whether the generated sequences are not only syntactically correct (as measured by NMT metrics) but also \textit{biologically functional}.} To evaluate binding specificity over a set of \(N\) generated pairs, we define the mean negative log-probability as
  \[
  \text{BS} = \frac{1}{N} \sum_{n=1}^{N} -\log \, p_{\text{bind}}(\hat{\mathbf{x}}^{(n)}, \hat{\mathbf{y}}^{(n)}).
  \]
  
  \item \textbf{Sequence Novelty}: Let \(\mathcal{X} = \{\hat{\mathbf{x}}_i\}_{i=1}^{M}\) denote the set of generated sequences and \(\mathcal{D}\) the training set. The novelty of a generated sequence is defined as the minimum edit distance to any sequence in \(\mathcal{D}\):
  \[
  \eta(\hat{\mathbf{x}}; \mathcal{D}) = \min_{\mathbf{x} \in \mathcal{D}} d(\hat{\mathbf{x}}, \mathbf{x}),
  \]
  where \(d(\cdot,\cdot)\) denotes the edit distance. The overall novelty metric is then given by the median of these minimum distances:
  \[
  \text{Novelty}(\mathcal{X}, \mathcal{D}) = \text{median}\left\{\eta(\hat{\mathbf{x}}; \mathcal{D}) : \hat{\mathbf{x}} \in \mathcal{X}\right\}.
  \]
  \textbf{This metric is essential for evaluating the model's \textit{de novo} design capabilities.}
\end{itemize}

In summary, these rigorously defined metrics provide a holistic evaluation framework for our bidirectional sequence generation approach, ensuring that our model achieves high \textbf{syntactic} quality and low uncertainty while remaining computationally efficient and biologically relevant.

\subsection{Bidirectional Training Framework}
Our training approach is designed to capture the dual nature of TCR-peptide recognition. In the immune system, TCRs evolve to recognize specific peptides, while certain peptides can be recognized by multiple TCRs. \textbf{As illustrated in Figure~\ref{fig:2},} we mirror this bidirectional relationship in our training framework.

The training process uses a technique called teacher forcing \cite{lamb2016professor}, where during training, we provide the model with both the input sequence and the correct (\textit{ground-truth}) output sequence. This approach helps the model learn the precise \textbf{mapping} relationships between TCRs and peptides \textbf{by providing the correct prefix at each timestep, which also effectively mitigates error propagation issues during the early stages of training}. \textbf{As defined in Section 3.1,} the overall training objective \textbf{is a symmetric loss function that} combines two directions:
\begin{equation}
    \mathcal{L}_{\text{total}}(\theta) = \mathcal{L}_{\text{TCR2PEP}}(\theta) + \mathcal{L}_{\text{PEP2TCR}}(\theta)
\end{equation}

For generating peptides from TCRs (TCR2PEP):
\begin{equation}
    \mathcal{L}_{\text{TCR2PEP}}(\theta) = -\sum_{n=1}^N \sum_{t=1}^{L_y} \log p_\theta(y_t^{(n)} \mid \mathbf{x}^{(n)}, \mathbf{y}_{<t}^{(n)})
\end{equation}

And for generating TCRs from peptides (PEP2TCR):
\begin{equation}
    \mathcal{L}_{\text{PEP2TCR}}(\theta) = -\sum_{n=1}^N \sum_{t=1}^{L_x} \log p_\theta(x_t^{(n)} \mid \mathbf{y}^{(n)}, \mathbf{x}_{<t}^{(n)})
\end{equation}

The combined loss functions represent the model's learning objective: maximizing the \textbf{joint} probability of generating correct sequences in both directions. We additionally implement gradient clipping (preventing too large updates) and dropout (randomly deactivating parts of the model during training to prevent over-reliance on specific patterns) to ensure stable training.
\begin{table*}[htbp]
\centering
\resizebox{\textwidth}{!}{%
\begin{tabular}{@{}llccccccccc@{}} 
\toprule
\multirow{2}{*}{\textbf{Model Type}} & \multirow{2}{*}{\textbf{Model}} & \multirow{2}{*}{\textbf{Pre-training}} & \multicolumn{4}{c}{\textbf{NMT Metrics}} & \multicolumn{2}{c}{\textbf{Efficiency}} & \multicolumn{2}{c}{\textbf{Bio Metric}}\\
\cmidrule(r){4-7} \cmidrule(l){8-9} \cmidrule(l){10-11}
& & & \textbf{BLEU} & \textbf{Edit Distance} & \textbf{F1 Score} & \textbf{Perplexity} & \textbf{Inf. Speed} & \textbf{Params} & \textbf{Binding Specificity} & \textbf{Novelty}\\
\midrule
\multirow{6}{*}{\textbf{TCR2PEP}} 
& LSTM & -      & 26.196 & 6.51 & 0.48 & 8.425  & 0.0053 & 7.38M  & 0.4515 & 5.0\\
& GRU  & -      & 32.022 & 5.97 & 0.52 & 79.035 & 0.0062 & 9.87M  & 0.4575 & 5.0\\
& BiLSTM & -    & 34.831 & 5.77 & 0.54 & 60.868 & 0.0058 & 13.16M & 0.4517 & 5.0\\
& CNN  & -      & 12.575 & 7.69 & 0.41 & 2.967  & 0.0129 & \textbf{2.39M} & 0.4505 & 5.0\\
& BART & +     & 13.085 & 8.05 & 0.40 & 2.616  & 0.0261 & 139M   & \textbf{0.4751} & \textbf{8.0}\\
& T5   & +     & \textbf{38.264} & \textbf{5.29} & \textbf{0.56} & \textbf{1.958} & 0.0225 & 223M   & 0.4397 & \textbf{8.0}\\
\midrule
\multirow{6}{*}{\textbf{PEP2TCR}} 
& LSTM & -      & 17.825 & 8.86 & 0.60 & 10.687 & 0.0074 & 7.38M  & 0.4620 & 2.0\\
& GRU  & -      & 18.227 & 8.93 & 0.61 & 14.589 & 0.0078 & 9.87M  & 0.4471 & 2.0\\
& BiLSTM & -    & 18.010 & 8.85 & 0.61 & 14.759 & 0.0076 & 13.16M & 0.4474 & 2.0\\
& CNN  & -      & 21.633 & 7.87 & 0.61 & 6.068  & 0.0176 & \textbf{2.39M} & 0.6400 & 1.0\\
& BART & +     & 23.412 & 7.55 & \textbf{0.64} & 2.425  & 0.0269 & 139M   & 0.5642 & \textbf{13.0}\\
& T5   & +     & \textbf{24.036} & \textbf{7.39} & 0.62 & \textbf{1.912} & 0.0311 & 223M   & \textbf{0.6618} & \textbf{13.0}\\
\bottomrule
\end{tabular}%
}
\caption{Evaluation results for TCR2PEP and PEP2TCR models using various architectures and decoding strategies. (``+'' indicates use of pre-trained parameters; ``Inf. Speed'' denotes inference speed in seconds per sequence.)}
\label{tab:results}
\end{table*}
\subsection{Baseline Methods and Decoding Strategies}
Our modeling framework evaluates a diverse array of sequence-to-sequence architectures that capture the complex, many-to-many mapping between TCR and peptide sequences. \textbf{We intentionally selected a spectrum of models spanning different inductive biases to establish a comprehensive benchmark:}
\begin{itemize}[noitemsep, leftmargin=*]
    \item \textbf{Classical Recurrent Models:} Including LSTM~\cite{hochreiter1997long}, GRU~\cite{chung2014empirical}, and BiLSTM~\cite{schuster1997bidirectional}. \textbf{These serve as standard baselines for sequential modeling.}
    \item \textbf{Autoregressive CNN Variants:} ~\cite{van2016pixel}. \textbf{This architecture exploits local receptive fields to identify position-invariant sequence motifs.}
    \item \textbf{Transformer-based Models:} Such as BART~\cite{lewis2019bart} and T5~\cite{raffel2020exploring}, which are fine-tuned for the TCR-peptide generation task. \textbf{These are used to evaluate the performance gains from large-scale pre-training and the attention mechanism.}
\end{itemize}
This diverse set of models is designed to evaluate the impact of architectural inductive biases and pre-training on the generation of novel sequences, ultimately aiding the \textit{de novo} design of antigenic peptides and potentially novel TCRs.

During the actual generation of sequences (inference phase), our model leverages the high-dimensional patterns learned from training on paired sequence data to generate one amino acid at a time. This sequential generation process is guided by two main principles:

\textbf{Principle 1: Autoregressive Generation:} The model generates each amino acid based on both the input sequence and all previously generated amino acids. For TCR2PEP:
\begin{equation}
    p_\theta(\mathbf{y} \mid \mathbf{x}) = \prod_{t=1}^{L_y} p_\theta(y_t \mid \mathbf{x}, \mathbf{y}_{<t})
\end{equation}

\textbf{Principle 2: Multiple Generation Strategies:} We implement four complementary approaches for sequence generation, each serving different aspects of protein sequence design:

    a) Greedy search: This strategy selects the amino acid with the highest probability at each step, providing focused exploration of \textbf{the model's most confident} predictions:
    \begin{equation}
        y_t = \arg\max_{y \in \mathcal{V}} p_\theta(y \mid \mathbf{x}, \mathbf{y}_{<t})
    \end{equation}

    b) Top-k sampling: By randomly sampling from the k most likely amino acids, this approach enables \textbf{controlled exploration} of sequence variations while maintaining prediction confidence within a specified range:
    \begin{equation}
        y_t \sim \text{Top-k}(p_\theta(y \mid \mathbf{x}, \mathbf{y}_{<t}))
    \end{equation}

    c) Top-p sampling (nucleus sampling): This method provides \textbf{adaptive sampling} based on the probability distribution at each position, naturally adjusting between conservative and diverse predictions depending on the underlying amino acid distribution:
    \begin{equation}
        y_t \sim \{y : \sum_{y' \leq y} p_\theta(y' \mid \mathbf{x}, \mathbf{y}_{<t}) \leq p\}
    \end{equation}

    d) Beam search: Through maintaining multiple candidate sequences simultaneously, this approach enables exploration of different high-probability sequence combinations, offering a balance between \textbf{local and global sequence} optimization:
    \begin{equation}
        \mathbf{y}^* = \arg\max_{\mathbf{y}} \sum_{t=1}^{L_y} \log p_\theta(y_t \mid \mathbf{x}, \mathbf{y}_{<t})
    \end{equation}

\par We incorporate a spectrum of sequence generation strategies, each playing a distinct yet complementary role. \textbf{Conservative decoding methods} (e.g., beam search and greedy search) provide high-fidelity predictions that align closely with known protein patterns, \textbf{which are crucial for optimizing the NMT metrics (see Section 3.3) such as BLEU}. In contrast, \textbf{exploratory methods} (e.g., top-k and top-p sampling) enable creative sequence discovery by systematically venturing beyond natural constraints, \textbf{which is essential for optimizing biological relevance metrics like Novelty}. This dual approach creates a powerful framework where conservative methods establish reliable foundations while sampling methods strategically expand the exploration space. Such a complementary strategy enables researchers to effectively navigate between preserving essential biological constraints and discovering novel protein candidates.

\begin{figure*}[h]
    \centering
    \includegraphics[width=0.95\textwidth]{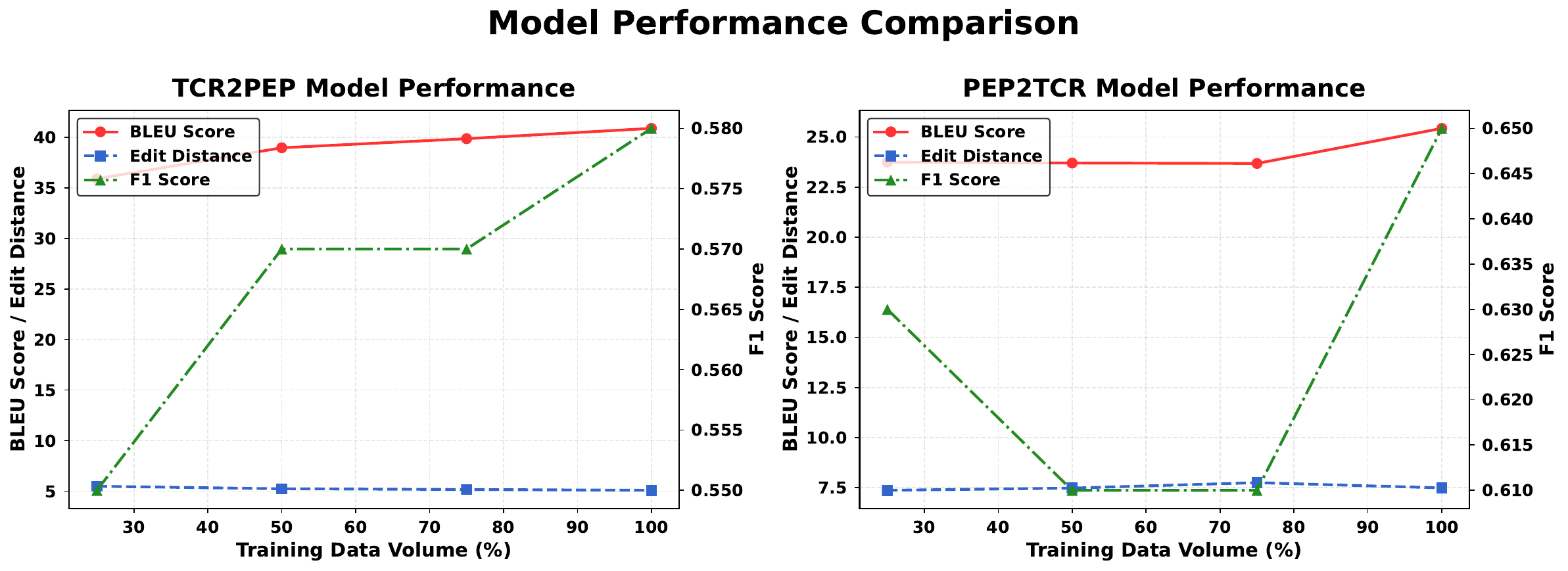} 
    \caption{Performance comparison under different training data volumes.}
    \label{fig:2}
\end{figure*}

\section{Results and Analysis} In this section, we first present our main benchmark experiments in Section 4.1, systematically comparing various neural network architectures—including LSTM, GRU, BiLSTM, CNN, BART, and T5—on the TCR2PEP and PEP2TCR sequence-generation tasks. Subsequently, based on the outcomes of these initial experiments and constrained by limited research time, we selected the top-performing T5 model to further explore the impact of model scaling, pre-training weights, and decoding strategies (Sections 4.2 and 4.3). These detailed analyses aim to provide practical insights and concrete guidelines for future model selection and optimization efforts in immunoinformatics research.

\subsection{Main Benchmark}

We conducted comprehensive experiments evaluating various neural architectures on the TCR2PEP and PEP2TCR tasks, specifically considering their applicability to protein sequence design. Results summarized in Table \ref{tab:results} demonstrate substantial differences across model architectures, pre-training strategies, efficiency metrics, and their implications for protein design.

\paragraph{Comparison of NMT Metrics.} For the TCR2PEP task, the T5 model, leveraging pre-trained parameters, achieves the highest BLEU (38.264), lowest Edit Distance (5.29), highest F1 Score (0.56), and lowest Perplexity (1.958). This significantly outperforms conventional recurrent architectures such as LSTM, GRU, and BiLSTM, suggesting that transformer-based models, especially those benefiting from large-scale pre-training, excel at capturing complex, long-range dependencies essential for accurate protein sequence prediction. CNN and BART models show notably weaker performance, implying that while pre-training is beneficial, architecture-specific mechanisms, such as attention in T5, greatly impact the ability to learn biologically meaningful sequence patterns.

In the reverse PEP2TCR task, T5 again yields optimal results, achieving the best BLEU (24.036), lowest Edit Distance (7.39), and lowest Perplexity (1.912), though BART slightly surpasses T5 in terms of F1 Score (0.64 vs. 0.62). The differences between T5 and BART here suggest distinct biological interpretations: BART's stronger F1 may reflect better recognition of biologically relevant subsequences or motifs, crucial in antigen recognition. However, T5’s overall robust performance indicates the value of capturing holistic sequence context provided by transformers.

\paragraph{Efficiency Analysis.} Considering inference speed and model size, CNN emerges as the most computationally efficient model (inference speed: 0.0129 s/sequence for TCR2PEP; 0.0176 s/sequence for PEP2TCR; 2.39M parameters). While its rapid performance is advantageous for high-throughput protein design experiments, CNN’s relatively weaker NMT performance limits its application to simpler design scenarios or preliminary screening processes. In contrast, T5 and BART, despite their superior accuracy and ability to capture biologically intricate relationships, exhibit significantly larger computational footprints (223M and 139M parameters, respectively). This highlights a trade-off critical for practical protein engineering pipelines, where computational cost and throughput must be balanced with sequence quality and biological relevance.

\begin{figure*}[h]
    \centering
    \includegraphics[width=0.9\textwidth]{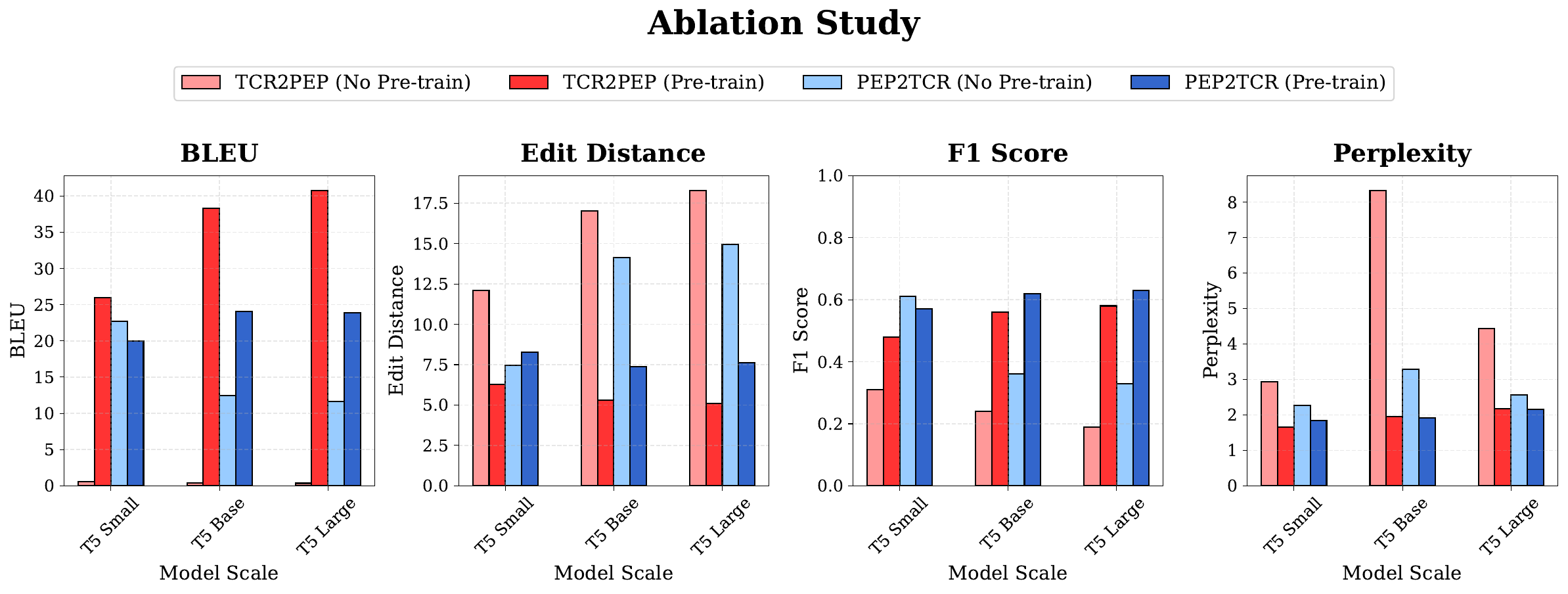} 
    \caption{Ablation study results for TCR2PEP and PEP2TCR models using T5 Small, Base, and Large architectures with and without pre-training.}
    \label{fig:2}
\end{figure*}

\paragraph{Bio Metric Evaluation.} Regarding biologically relevant metrics—Binding Specificity and Novelty—pre-trained transformers such as T5 and BART significantly outperform simpler architectures, highlighting the benefit of leveraging large-scale learned representations for biologically complex sequence design tasks. Specifically, for TCR2PEP, BART achieves the highest Binding Specificity (0.4751) and Novelty (8.0), which suggests superior ability in generating novel peptides with targeted binding properties, potentially due to BART’s nuanced internal representations fine-tuned from pre-training on diverse sequence contexts. For PEP2TCR, T5 exhibits the highest Binding Specificity (0.6618) and Novelty (13.0), indicative of an excellent capability in producing biologically viable and highly specific TCR sequences.

Interestingly, high performance on traditional NMT metrics does not strictly correlate with high Binding Specificity, highlighting a critical biological nuance. Specifically, sequences optimized solely for computational metrics like BLEU or Edit Distance may fail to adequately represent molecular-level interactions and structural constraints fundamental to protein function, such as precise antigen-TCR binding interfaces or critical residues essential for binding affinity. Thus, this decoupling emphasizes the necessity of integrating biological constraints directly into model training to ensure the functional relevance and biological utility of designed proteins. For example, in the TCR2PEP task, T5, despite superior general sequence generation metrics, demonstrates lower Binding Specificity than BART. This decoupling reveals an important biological insight: sequences that score highly in traditional computational metrics may not necessarily capture the intricate molecular interactions required for specific peptide-TCR recognition. Consequently, models must integrate biologically-driven training objectives or constraints explicitly designed to enhance molecular specificity.

Overall, our findings highlight critical considerations in selecting neural architectures for protein sequence design, emphasizing the importance of balancing computational efficiency, sequence accuracy, and biological relevance. Future work should further explore training methods explicitly tailored towards biologically-relevant optimization criteria, thereby enhancing the practical and translational value of machine learning models in protein engineering.

\subsection{Ablation on Scale and Pretraining Weight (T5)} Given the strong performance of T5 in our main experiments (Section 4.1), we further investigated the effects of scaling and pre-training on the TCR2PEP and PEP2TCR tasks using T5 models of different sizes (Small, Base, Large), with and without pre-training (Figure~\ref{fig:2}). Our experiments reveal three key findings: (1) Pre-training consistently boosts performance, notably improving BLEU scores in TCR2PEP; (2) larger architectures benefit disproportionately from pre-trained weights, indicating a synergistic effect between scale and pre-training; and (3) T5-Large with pre-training achieves the best edit distance and perplexity scores, confirming the value of large-scale pre-training for both sequence quality and generalization. These results suggest that leveraging larger, pre-trained models can effectively balance computational efficiency and prediction accuracy in practical immunoinformatics applications.

\subsection{Ablation on Decoding Strategy (T5-Large)} Focusing on T5-Large from our primary experiments (Section 4.1), we explored various decoding strategies by comparing greedy search, Top-k sampling ($k=10,30,50$), Top-p sampling ($p=0.7,0.8,0.9$), and beam search ($\text{beam}=2,4,8$) (Table~\ref{tab:decoding-ablation}). Three insights emerged: (1) Beam search, particularly with a beam size of 2, consistently yields the best performance across both tasks in terms of BLEU, edit distance, and F1 scores; (2) although stochastic methods like Top-k and Top-p sampling can enhance diversity, they tend to reduce precision in these biological tasks; and (3) TCR2PEP is relatively insensitive to beam size variations, whereas PEP2TCR shows performance declines with larger beam sizes. These findings provide practical guidance for selecting decoding strategies tailored to the unique demands of biological sequence generation.

\begin{table}[t!]  
\centering
\resizebox{1.0\linewidth}{!}{%
\begin{tabular}{@{}llccc@{}} 
\toprule
\textbf{Model Type} & \textbf{Decoding Strategy} & \textbf{BLEU} & \textbf{Edit Distance} & \textbf{F1 Score}\\
\midrule
\multirow{7}{*}{\textbf{T5-TCR2PEP}} 
& Greedy & 40.77 & 5.10 & 0.58\\
& Top-k (k=10) & 29.85 & 6.20 & 0.51\\
& Top-k (k=30) & 29.06 & 6.27 & 0.50\\
& Top-k (k=50) & 28.45 & 6.33 & 0.50\\
& Top-p (p=0.7) & 34.78 & 5.72 & 0.54\\
& Top-p (p=0.8) & 32.32 & 5.95 & 0.53\\
& Top-p (p=0.9) & 31.40 & 6.06 & 0.52\\
& Beam Search (Beam=2) & \textbf{40.87} & \textbf{5.09} & \textbf{0.58}\\
& Beam Search (Beam=4) & 40.86 & \textbf{5.09} & \textbf{0.58}\\
& Beam Search (Beam=8) & 40.86 & \textbf{5.09} & \textbf{0.58}\\
\midrule
\multirow{10}{*}{\textbf{T5-PEP2TCR}} 
& Greedy & 23.86 & 7.61 & 0.63\\
& Top-k (k=10) & 19.19 & 8.70 & 0.62\\
& Top-k (k=30) & 18.59 & 8.86 & 0.61\\
& Top-k (k=50) & 18.68 & 8.87 & 0.61\\
& Top-p (p=0.7) & 22.20 & 8.26 & 0.63\\
& Top-p (p=0.8) & 21.79 & 8.36 & 0.63\\
& Top-p (p=0.9) & 20.86 & 8.50 & 0.62\\
& Beam Search (Beam=2) & \textbf{25.43} & \textbf{7.49} & \textbf{0.65}\\
& Beam Search (Beam=4) & 25.10 & 7.53 & 0.65\\
& Beam Search (Beam=8) & 23.65 & 8.13 & 0.63\\
\bottomrule
\end{tabular}%
}
\caption{Ablation study of different decoding strategies for the T5-large model on TCR2PEP and PEP2TCR tasks.}
\label{tab:decoding-ablation}
\end{table}

\section{Discussion} Our comprehensive benchmark study introduces TCRTransBench, a systematic evaluation framework tailored for bidirectional TCR--peptide sequence generation, an essential task in computational immunology and therapeutic protein design. Unlike conventional TCR--peptide studies that mainly formulate the problem as discriminative binding prediction, TCRTransBench explicitly focuses on generative sequence-to-sequence modeling in both directions: generating antigenic peptides from TCR sequences and generating TCR sequences from antigenic peptides. This bidirectional formulation is particularly important for \textit{de novo} immune design, because practical therapeutic scenarios often require not only predicting whether a known pair binds, but also proposing novel candidate peptides or receptors with desired specificity.

By rigorously assessing recurrent, convolutional, and transformer-based architectures across multiple evaluation dimensions, our results reveal clear trade-offs between computational efficiency, sequence accuracy, and biological relevance. Transformer-based models, especially those initialized with pre-trained weights, generally achieve stronger sequence-generation performance, suggesting that attention-based architectures are better suited for capturing long-range dependencies and complex residue-level patterns in TCR--peptide interactions. However, this performance gain comes with substantially higher computational cost. In contrast, recurrent and convolutional models remain attractive under resource-constrained or high-throughput screening settings, where fast inference and smaller model size may be more important than achieving the best sequence-level accuracy. These findings indicate that model selection should be application-dependent rather than based on a single metric: large transformers are preferable for high-quality candidate generation, whereas lightweight models may serve as efficient preliminary filters.

A particularly important observation is the divergence between traditional NMT metrics and biologically relevant metrics. Metrics such as BLEU, edit distance, F1 score, and perplexity mainly evaluate syntactic similarity between generated and reference sequences. While these measures are useful for quantifying sequence-level fidelity, they do not necessarily reflect functional binding properties or biological plausibility. For example, a generated sequence may closely resemble a reference peptide at the character level but still fail to preserve key residues involved in TCR recognition. Conversely, a biologically plausible and novel sequence may receive a lower score under conventional text-generation metrics if it deviates from the reference sequence. This discrepancy highlights a central challenge in immunological sequence generation: optimizing for textual similarity alone is insufficient for therapeutic design. Future models should therefore incorporate biologically informed objectives, such as binding specificity, motif preservation, structural compatibility, and sequence novelty, either during training or decoding.

Our decoding strategy analysis further supports this conclusion. Conservative decoding methods, such as greedy search and beam search, tend to improve sequence accuracy by selecting high-probability outputs, making them suitable when fidelity to the training distribution is desired. However, these methods may reduce diversity and limit exploration of novel sequence space. In contrast, stochastic decoding strategies such as top-$k$ and top-$p$ sampling encourage sequence diversity, but may sacrifice precision and biological reliability. This trade-off suggests that controllable generation is essential for TCR--peptide design: an ideal system should allow users to adjust the balance between accuracy, diversity, and biological plausibility according to downstream experimental goals.

Overall, TCRTransBench provides a robust dataset, standardized bidirectional generation tasks, and a comprehensive evaluation protocol for future immunoinformatics research. More importantly, our results demonstrate that progress in this area should not be measured solely by improvements in conventional sequence-generation metrics. Instead, future work should move toward biologically grounded generative modeling, integrating functional binding constraints, structural priors, MHC presentation information, and experimental validation. Such extensions will be critical for transforming TCR--peptide sequence generation from a computational benchmark task into a practically useful tool for immunotherapy, vaccine design, and therapeutic protein engineering.

\clearpage

\begin{figure*}[p!]
    \centering
    \vspace*{\fill}
    \includegraphics[
        width=\textwidth,
        height=0.85\textheight,
        keepaspectratio
    ]{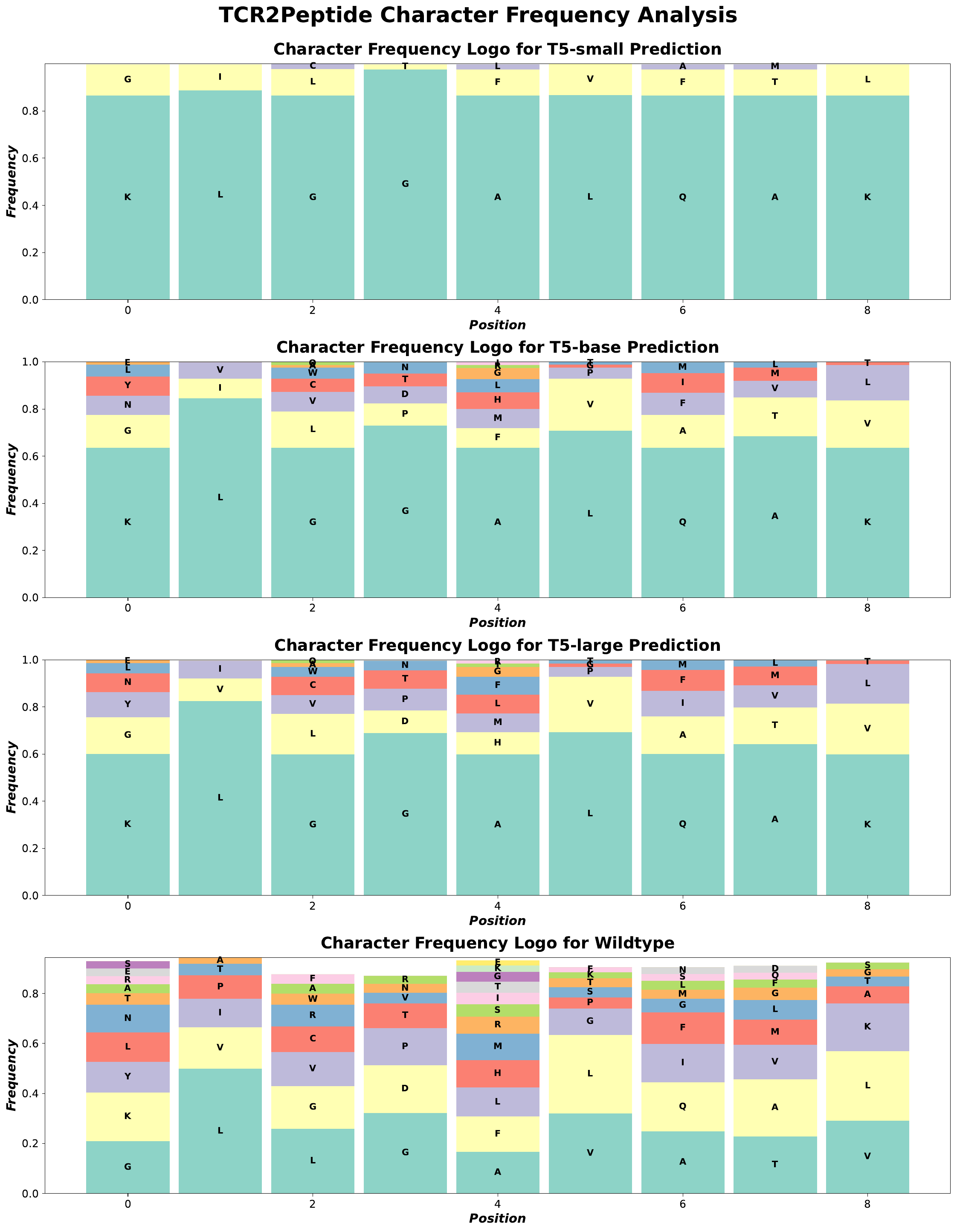}
    \caption{Position-wise amino acid frequency analysis for TCR2PEP predictions.}
    \label{fig:tcr2pep_frequency}
    \vspace*{\fill}
\end{figure*}

\clearpage

\section{Conclusion} In this work, we presented TCRTransBench, a systematic benchmark framework tailored for bidirectional TCR-peptide sequence generation tasks, which are crucial for advancing computational protein design in immunotherapy and vaccine development. Through extensive evaluation of diverse neural architectures and decoding strategies, our study elucidates the inherent trade-offs between sequence accuracy, computational efficiency, and biological relevance, and underscores the importance of incorporating biologically-informed metrics in model assessment. The insights gleaned from our benchmark provide a solid foundation for future investigations, ultimately guiding the development of more effective and specialized models in the rapidly evolving field of immunoinformatics.

\section{Limitations and Future Work}
Although the TCRTransBench framework provides a comprehensive evaluation for bidirectional TCR-peptide sequence generation, several limitations remain. 
First, the benchmark dataset, while rigorously curated, is restricted to publicly available human TCR-peptide pairs, which might limit its representativeness and generalizability. Future efforts should extend dataset diversity and coverage, including additional species or broader immunological contexts. 
Second, our baseline methods, though representative, are not exhaustive; exploring other advanced modeling paradigms could further enhance performance, and we encourage the community to develop and benchmark novel computational approaches tailored explicitly for these tasks. 
Third, our current generation framework operates in an MHC-free manner. While this enables broader exploration of the TCR-peptide sequence space, it does not explicitly account for MHC presentation rules, such as antigen processing and MHC binding stability. Consequently, some generated peptides, despite having high predicted TCR affinity, might not be effectively presented by specific HLA alleles in vivo.
Finally, evaluations currently rely on computational predictions such as the Pan-Peptide model, whose accuracy may not fully reflect biological realities. Incorporating experimental validation through laboratory assays remains crucial to reliably assess the functional validity of generated sequences in future studies.

\bibliography{example_paper}
\bibliographystyle{icml2025}

\appendix

\end{document}